\documentclass{sig-alternate-05-2015}

\begin{document}

\setcopyright{acmcopyright}


\isbn{123-4567-24-567/08/06}

\acmPrice{\$15.00}

%

\title{From Query-By-Keyword to Query-By-Example: LinkedIn Talent Search Approach}

%
%
%
%
%

\numberofauthors{1} 
%
\author{
%
%
\alignauthor
{Viet Ha-Thuc, Yan~Yan, Xianren~Wu, Vijay~Dialani, Abhishek~Gupta, Shakti~Sinha}\\
\affaddr{LinkedIn Corporation\\ 2029 Steirlin Ct, Mountain View, CA 95035, USA}\\
\email{vhathuc@fb.com, \{cyan,rwu,vdialani,agupta,ssinha\}@linkedin.com}
}

\maketitle
\begin{abstract}
One key challenge in talent search is to translate complex criteria of a hiring position into a search query, while it is relatively easy for a searcher to list examples of suitable candidates for a given position.  To improve search efficiency, we propose the next generation of talent search at LinkedIn, also referred to as  Search By Ideal Candidates. In this  system, a searcher provides one or several ideal candidates as the input to hire for a given position. The system then  generates a query based on the ideal candidates and uses it to retrieve and rank results. Shifting from the traditional Query-By-Keyword to this new Query-By-Example system poses a number of  challenges:  How to generate a query that best describes the candidates? When moving to a completely different paradigm, how does one leverage previous product logs to learn ranking models and/or evaluate the new system with no existing usage logs? Finally, given the different nature between the two search paradigms, the ranking features typically used for Query-By-Keyword systems might not be optimal for Query-By-Example. This paper describes our approach to solving  these challenges. We   present experimental results confirming the effectiveness of the proposed solution, particularly on query building and search ranking tasks. As of writing this paper, the new system has been available  to all LinkedIn members.
\end{abstract}

%
%

%
%

%
%
\printccsdesc



\section{Introduction}
LinkedIn is the largest professional network on the Internet with more than 450 million members. Its vision is to create the world's first economic graph matching talent and opportunity at a massive scale. Over time, as its member base increases and its economic graph vision becomes reality, LinkedIn has become the source for corporations around the world to find new employees. On the financial side, about {\it 64\%} of the company revenue is from Talent Solutions\footnote{https://press.linkedin.com/about-linkedin}, which is a product helping recruiters and corporations around the world search for talent. Therefore, talent search problem is extremely important to LinkedIn both in terms of value propositions it provides to its members as well as its revenue stream.

On one hand, the main challenge in talent search is to translate the criteria of a hiring position into a search query that leads to desired candidates. To achieve this goal, the searcher has to understand which skills are typically required for a given position, what the alternatives are, which companies are likely to have such candidates, which schools the candidates are most likely to graduate from, etc. Moreover, such knowledge changes over time. As a result, it is not surprising to find that even for experienced recruiters, it often requires many search trials for formulating a good query, as observed in the LinkedIn search log data. 

Alternatively, it is usually easier for a searcher to pick one or several examples of good candidates for a given position. For instance, hiring managers or recruiters can simply source by using  existing team members as ideal candidates. Motivated by such use cases, we propose a new talent search approach called \textit{Search By Ideal Candidates}. In this new approach, instead of specifying a complex query capturing the position requirements, the searcher provides  a small set of ideal candidates (typically from one to three candidates) suitable for the position. The system then builds a query from the input candidates and provides the query to the searcher. It is worth emphasizing that the query built from the ideal candidates is not only an intermediate goal used to find result candidates but something a searcher could continue to modify. Exhibiting the query allows the searcher to rationalize why a certain result is shown up in search results thus making  the system transparent to the searcher. Subsequently, the  system retrieves the candidates matching the query and ranks them with respect to the query as well as the ideal candidates.

However, moving from the traditional Query-By-Keyword to Query-By-Example paradigm raises a set of   technical challenges. The first challenge is how to extract the key information from each of the ideal candidates' profiles and aggregate them to generate a query best describing the information need that the candidates collectively represent. Second, since talent search is highly personalized, the most typical way to get training data to learn and evaluate ranking functions is from usage logs \cite{hathuc2015expertisesearch}. With no existing logs for the new paradigm, we have to invent ways of repurposing the logs from the Search-By-Keyword paradigm. The final open question is, given the difference between the two search paradigms, which new signals we should add to the ranking function to make it suitable for a Query-By-Example system. To the best of our knowledge, there is no prior work comprehensively addressing these challenges in an industrial setting.

In this paper, for the query building challenge, we propose an approach that in addition to taking  the ideal candidates' profiles into account, also leverages a large volume of semi-structured data and social affiliations on LinkedIn professional network. This approach filters out outliers in the profiles, extracts the relevant information from the profiles, and  infers missing relevant information  to generate queries. To overcome the lack of usage logs of the new \emph{Search By Ideal Candidates} system, we propose a generative model to infer labeled data for \emph{Search By Ideal Candidates} from usage logs of the previous Query-By-Keyword system. Even though our work is rooted in  talent search,  we believe the approach is also useful when developing a Query-By-Example from Query-By-Keyword system in other domains. Finally, in order to adapt the search ranking, we introduce a new set of features that directly capture the resemblance between the ideal candidates and the retrieved results on various aspects including expertise, job title, career-trajectory similarities.   

Our experimental results confirm the effectiveness of the proposed query building and search ranking approaches. In particular, when evaluated on a randomized test set collected from live traffic, the new ranking model is $13.1\%$ better in terms of NDCG@5 than the ranking model currently used in the Query-By-Keyword system. When compared with a second baseline that is trained specifically for Query-By-Example use case but does not use the proposed features, the new model is still $6.2\%$ better. That demonstrates the benefit of the aforementioned new feature set.

The rest of the paper will be organized as follows. Section 2 reviews related work. Section 3 presents an overview of the whole system. Sections 4 and 5 discuss the two most important steps: query building and learning ranking function. Experimental results are shown in Section 6. Finally, we conclude our works in Section 7.

Disclaimer: The first version of \emph{Search By Ideal Candidates} system was previously demonstrated at WWW 2016 \cite{hathuc2016talentsearch}. However, in this paper, we rigorously focus on technical aspects of the work instead of system demonstration. Moreover, this work also presents significant improvements recently made. Specifically, we propose an approach to generating training data as well as novel features specifically designed for \emph{Search By Ideal Candidates} paradigm. Finally, we also conduct analyses to give deeper insights on the system effectiveness as well as detail various design tradeoffs and practical lessons learned during the course of the work.   

\section{Related Work}
Our work in this paper is related to several previous research directions in the literature including: (i) Query-By-Example; (ii) relevance feedback in text retrieval; (iii) item-to-item recommendation; (iv) learning to rank. In this section, we review these directions and differentiate our work from previous ones in each direction.
\subsection{Query By Example}
\label{subsec:query_by_example}
Query-By-Example ({QBE}) in image retrieval (also referred as content-based image retrieval), especially in dermatology studies, originates  from the seminal work in \cite{chang1984picture,rui1998relevance} in 1980s and 1990s. QBE  in image retrieval systems typically utilizes the example images containing objects of interest to generate visual features which include color and texture for  detecting and identifying the particular objects easily. More recently, there has been a focus on Query-By-Example in text domain \cite{YangBDIKP09, WengLCZZYZ11}. Text based QBE  systems extract key phrases from unstructured query documents by using techniques like tf-idf and also combine them with semantic matching, such as, LSI and LDA.    

Although conceptually  similar, the previous work and {QBE} in talent search differ significantly.  Firstly, from an  entity understanding aspect, {QBE} in image retrieval and text retrieval focuses on semantically simple features, whereas, talent search relies  on entity understanding for richer  features that capture interaction effects between entities.  For example, people's professional signatures (e.g., career trajectory, endorsed skills and social connections) are hierarchically structured and contain  complex  information compared to images or text documents. Secondly, from a retrieval system's perspective, the goal of {QBE} for image retrieval is to justify the existence of certain objects within an image.  However, in talent search, the goal is to accurately capture the searchers intent. Thus, talent search not only relies on the examples themselves but also the search context to precisely capture what exact skills, experience, etc., that the searchers are seeking  and other perspectives that are contextual to their use of the system.

\subsection{Item-To-Item Recommendation}
\label{subsec:item_to_item_recommendation}
Since the seminal work of Sarwar et al. \cite{sarwar2001item} and Balabanovi\'{c} et al. \cite{balabanovic1997fab}, item-based collaborative filtering has been widely used to identify similarity between recommended items. Later efforts include, but are not limited to \cite{schein2002methods,linden2003amazon,wang2006unifying,koren2008factorization}. \cite{wang2006unifying} provide probabilistic frameworks which make the collaborative filtering more robust when dealing with sparse data.  Schein et al. \cite{schein2002methods} combine the content-based approach (Section \ref{subsec:query_by_example}) and collaborative filtering together so that the whole approach could handle the cold-start problem. In \cite{koren2008factorization}, Koren combines collaborative filtering with latent factors to  utilize both the neighborhood information and latent factors. In our own previous work \cite{xu14}, we use such a technique to define  career-path similarity for similar people recommendation system on LinkedIn. For a comprehensive coverage of the topic, please refer to the survey \cite{su2009survey}. 

Unlike the item recommendation problem, our case, i.e., \emph{Search by Ideal Candidates} has a strong emphasis on making the recommendations ``explainable'' -  it is not only important to generate relevant results but also critical to build descriptive queries. These queries are also presented to searchers in order to provide transparency on  why a certain result is shown up in search ranking. Moreover, this allows the searchers to have control on search results by interacting with the presented queries, which is noted as query edit and rewrite. Therefore, our case uniquely combines aspects of search and recommender systems.

\subsection{Relevance Feedback in Information \\ Retrieval}
Relevance feedback is a popular approach used in text retrieval. It leverages searcher's feedback on an initial set of retrieval results to refine the original query. The query refinement could be in the form of re-weighting the query terms or automatically expanding the query with new terms. Rocchio \cite{rocchio1971relevance} is widely  considered to be  the first formalization of relevance feedback technique, developed on the vector space model. He proposes  query refinement based on the difference between the average vector of the relevant documents and the average vector of the non-relevant documents. A subsequent approach uses probabilistic models that estimate the probability a document is relevant to the information need \cite{RobertsonJ76,LavrenkoC01}, where the probabilistic models are inferred from the feedback documents.

Unlike the previous work focusing on free text, both input documents (ideal candidates) and generated queries in our case are semi-structured. Each field in the documents and queries has unique characteristics and requires a different query building strategy. Moreover, the ideal candidates (examples in the QBE) are also used to construct various features in the machine-learnt ranking function.

\subsection{Learning to Rank}
Learning a ranking function for \emph{Search By Ideal Candidates}, that incorporates signals from examples in QBE has shown promising results in our work. Learning to rank (LTR) has been central to information retrieval (IR),  as most systems typically use many features for ranking and that makes it difficult to manually tune the ranking functions.  There has been a lot of research published in the area of LTR, typically categorized  into one of the three following categories: (a) pointwise approaches view ranking as traditional binary classification or regression problems. (b) pairwise approaches take input as a set of pairs of documents in the form of one document is more relevant than the other with respect to a specific query. These approaches then learn a ranking function minimizing the number of incorrectly ordered pairs. (c) the state-of-the-art for LTR is listwise approach. This approach typically views the entire ranked list of documents as a learning instance while optimizing some objective function defined over all of the documents, such as, normalized discounted cumulative gain (NDCG) \cite{Cao2007}. We refer the readers who are interested in more details of LTR to~\cite{Chapelle2011b, Liu2010, Li2011} for more comprehensive reviews. 
 
A key element of LTR is collecting ground truth labeled data. Traditionally, ground truth data is labeled by either professional editors or crowd sourcing~\cite{Chapelle2011b}. However, the issues with this approach are (i) it is expensive and not scalable and (ii) it is very hard for the judges to evaluate the relevance on behalf of other users, making it challenging to apply the approach for personalized ranking. For these reasons, some previous research proposes to extract labeled data using click logs~\cite{Joachims2002, Craswell2008}.

However, unlike the previous work, when moving from Query-By-Keyword to Query-By-Example, the new system does not have logs to learn from. Instead, we propose an approach to generating training data for the new search paradigm by leveraging the logs of the traditional searching scheme.

\section{System Overview}
\begin{figure}
\centering
\includegraphics[width=0.33\textwidth]{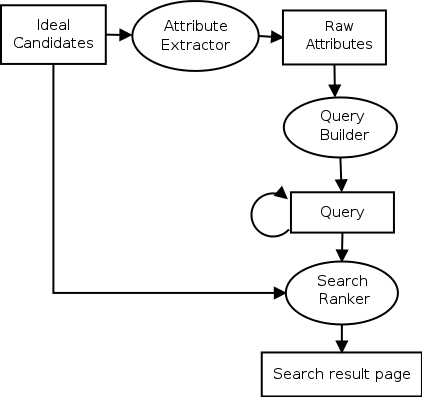}
\caption{System Overview}
\label{system_overview}
\end{figure}

Given a set of input ideal candidates selected by a searcher, the system builds a search query by capturing key information in their profiles and utilizes  the query to retrieve and rank results. The overall flow is shown in Figure \ref{system_overview}. In the first step, we extract the raw attributes, including skills, companies, titles, schools, industries, etc. from the ideal candidates' profiles individually. These raw attributes are then passed to a query builder. For each attribute type, the query builder aggregates the raw attributes across the input candidates, expands them to similar attributes and finally selects the top ones that best represent the ideal candidates. 

Once the query builder generates a query, the query is shown to the searcher and used to retrieve  results that are similar to the ideal candidates. The searcher can interact and modify the query, for instance, adding or removing some attributes like skills, titles or companies in the query. After the query is modified, search ranker refreshes the search results. Unlike the traditional search ranking functions, such as \cite{Liu2010,hathuc2015expertisesearch}, that are functions of query(-ies), result(s) and searcher(s) (if personalized), the search ranking function in our system takes all of them as well as the input ideal candidates into account. In the next sections, we discuss in more details how to generate a query from input ideal candidates and how to learn a search ranking function specifically for \emph{Search By Ideal Candidates}.

\section{Query Building}
\label{query_building_sec}
\begin{figure}
\centering
\includegraphics[width=0.48\textwidth]{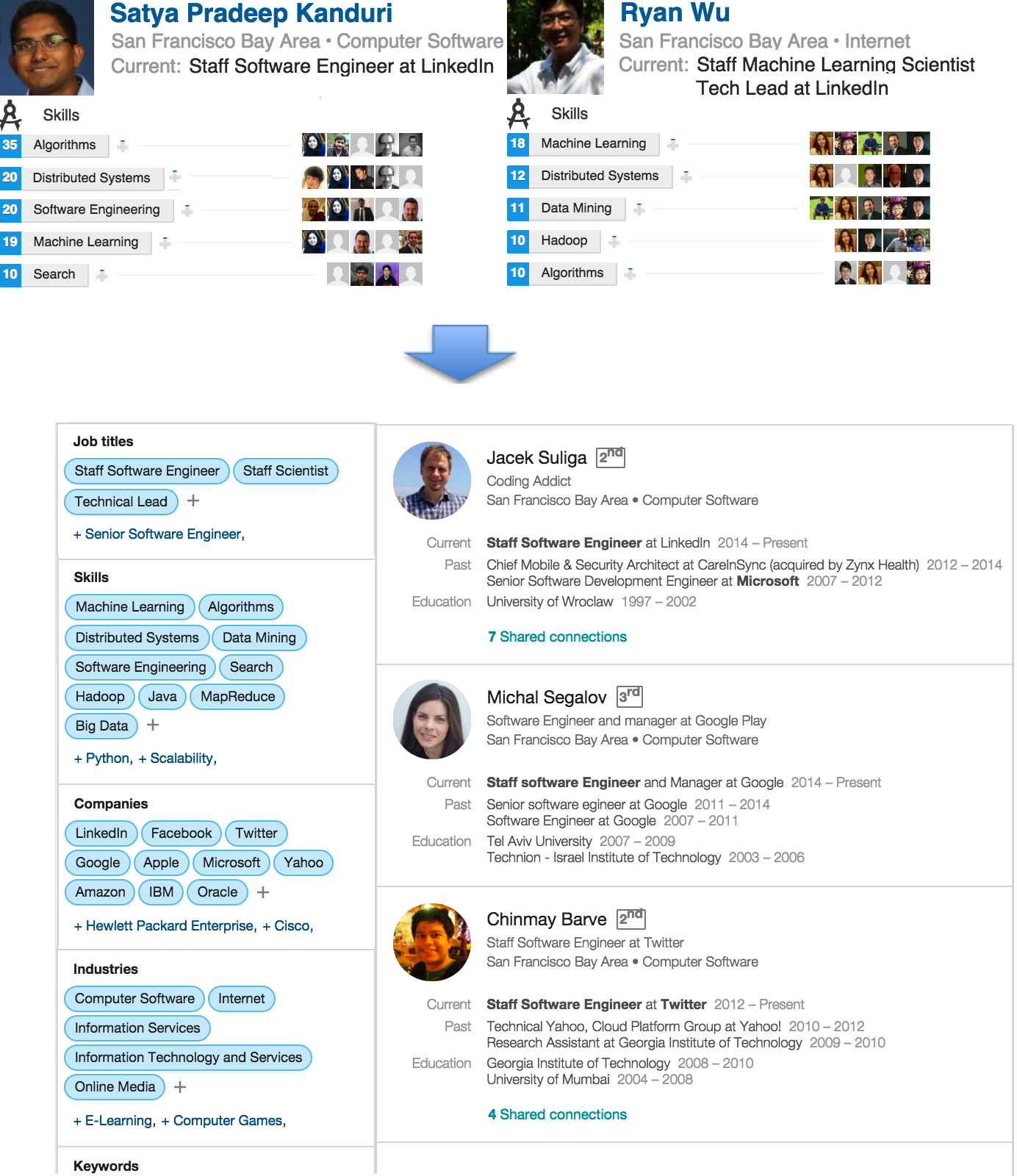}
\caption{Query Building}
\label{query_building}
\end{figure}

After a searcher selects one or a few ideal candidates, the query builder generates a query that contains titles, skills, companies and industries as demonstrated in Figure \ref{query_building}. The upper part of the figure shows snippets of the two selected candidates' profiles. The lower part shows the GUI of the system. The generated query (on the left rail) contains values for each of the four attribute types. Skill attribute includes ``machine learning'', ``Algorithm'', ``Distributed Systems", etc. These are the skills that the query builder believes best represent expertise of the candidates. Similarly, job title, company and industry fields contain the corresponding entities that results similar to the ideal candidates are likely to have. In our system, different variations, such as, ``tech lead'' and ``technical lead'' are mapped (standardized) to the same entity. After observing the query, the searcher can delete entities that he thinks irrelevant. He can also add new ones by either selecting the ones that the system suggests (e.g., ``Senior Software Engineer'' in the figure) or manually entering them.  

Given the query displayed on the UI, the actual search query sent the backend search engine is simply a conjunction across the attribute types. In each attribute type, we take a disjunction across the selected entities. For instance, the search query for the example in Figure \ref{query_building} would be: \textit{Q = title:(``Staff Software Engineer'' OR ``Staff Scientist'' OR ``Technical Lead'') AND skill:(``Machine Learning'' OR ... OR ``Big Data'') AND company:(``LinkedIn'' OR ... OR ``Oracle'') AND industry:(``Computer Software'' OR ... OR ``Online Media'')}. This query is used to retrieve and rank the results, which are shown on the right rail in the figure. Due to space limit, in this paper, we focus on how to generate skill facet, which is usually the most informative part of a query.

\subsection{Skill Selection}
LinkedIn allows members to add  {\it skills} to their profiles. Typical examples of skills for an information retrieval researcher would be ``search'', ``information retrieval'', ``machine learning'', etc. On LinkedIn, there are more than 35 thousand standardized skills. Members can also {\it endorse} skills of other members in their network. Thus, skills are an integral part of members' profiles that showcases their professional expertise. A challenge is that the ideal candidates may not explicitly list all the skills they have on profiles. On the other hand, some of their skills might not be relevant to their core expertise. For instance, an information retrieval researcher could have ``nonprofit fundraising'' skill.

\begin{figure}
\centering
\includegraphics[width=0.3\textwidth]{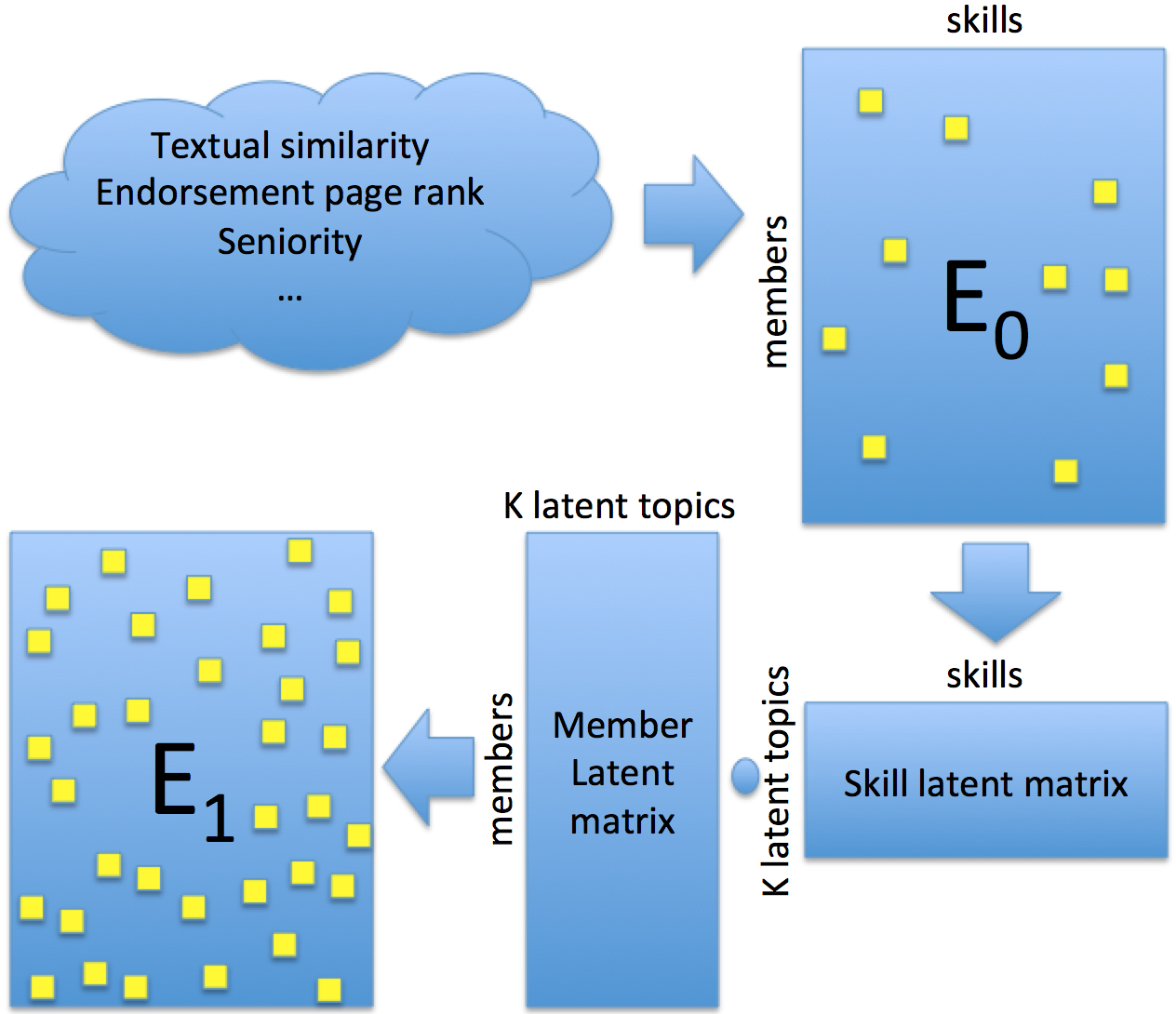}
\caption{Member-skill expertise estimation}
\label{member_expertise}
\end{figure}

To overcome these challenges, we first estimate expertise scores of a member on the explicit skills and the ones he might have. Figure \ref{member_expertise} describes the offline process to estimate the expertise scores. In the first step, we use a supervised learning algorithm combining various signals on LinkedIn such as skill-endorsement graph page rank, skill-profile textual similarity, member's seniority, etc. to estimate the expertise score, i.e., $p(expert | member, skill)$. After this step, the expertise matrix ($E_0$) is very sparse since we can be certain only for a small percentage of the pairs. In the second step, we factorize the matrix into member and skill matrices in K-dimensional latent space. Then, we compute the dot-product of the matrices to estimate the ``unknown'' cells. The intuition is that if a member has ``machine learning'' and ``information retrieval'' skills, based on skill co-occurrence patterns from all of our member base, we could infer that the member is also likely to know ``learning-to-rank''. Since the dot-product results in a large number of non-zero scores of each member on the 35K skills, the scores are then thresholded. If member's score on a skill is less than a threshold, the member is assumed not to know the skill and has that expertise score corresponding to zero. Thus, the final expertise matrix ($E_1$) is still sparse but much denser than $E_0$. We refer interested readers to our recent work \cite{hathuc2015expertisesearch} for more details. 

\begin{equation}
	f(sk) = \sum_{c \in IC} {expertiseScore(c, sk)}
\label{skill_ranking}
\end{equation}

At runtime, given a set of input ideal candidates $IC$, we rank the skills by Equation \ref{skill_ranking}. The top-N skills are then selected to represent the ideal candidates. Expertise scores of an ideal candidate on outlier skills are typically very low, resulting in unselected. Moreover, by taking the sum over all candidates, we boost the skills that many candidate profiles contain, to represent the commonality of the skill set.

\section{Learning To Rank for Search-By-Ideal-Candidates}
\label{ltr_sbic_sec}
\subsection{Training Data Generation}
\label{training_data_generation}
\begin{figure}
\centering
\includegraphics[width=0.3\textwidth]{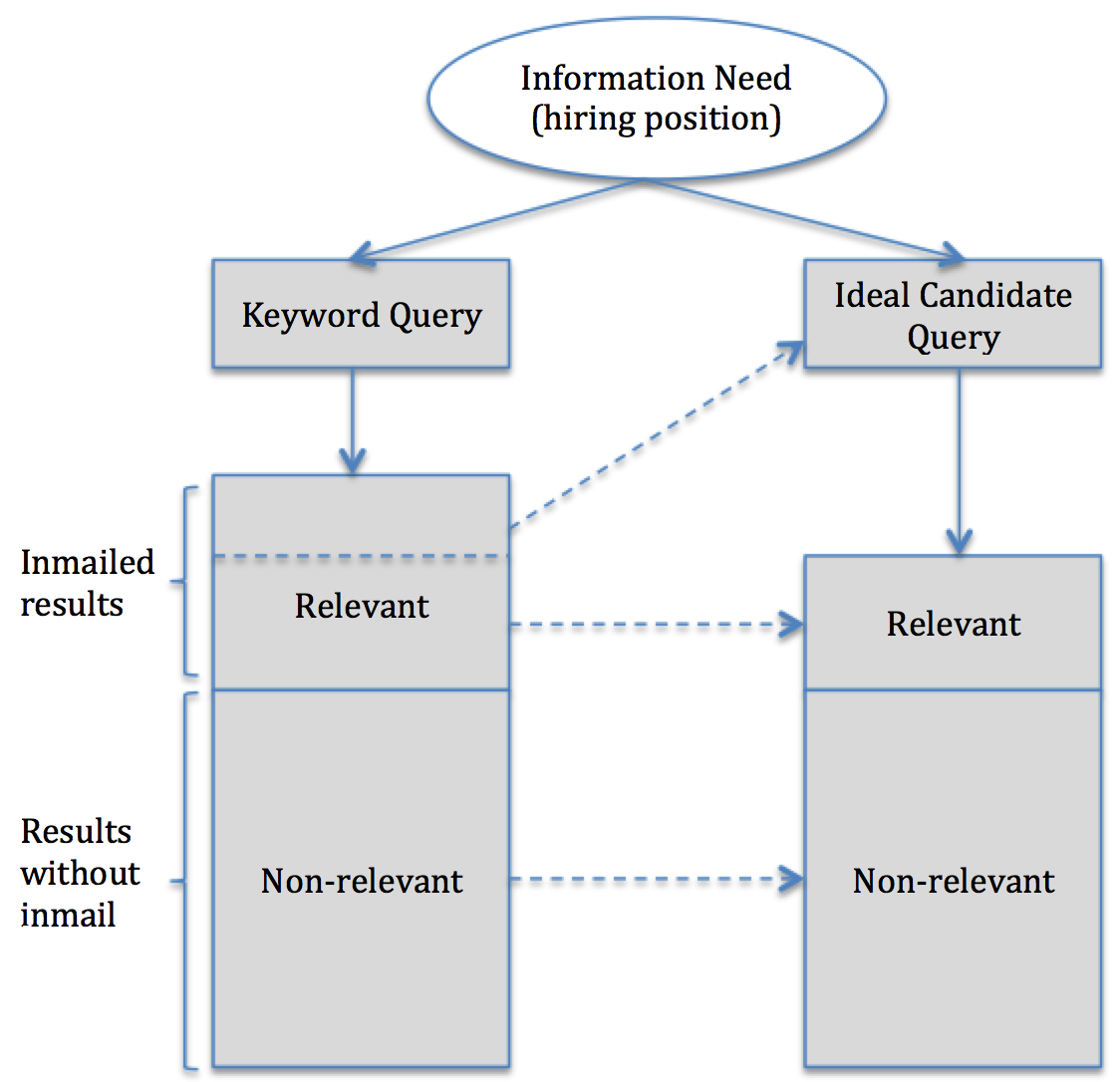}
\caption{Generate training data for Search By Ideal Candidates from usage logs of a Query-By-Keyword system}
\label{co-inmailed}
\end{figure}

As mentioned before, since \emph{Search By Ideal Candidates} is a brand new product, it does not have any usage logs from which we can generate training data at the time the system is being built. To overcome this, we propose a generative model to infer training data from the logs of our previous Query-By-Keyword system. In this generative model, an information need (hiring position in searcher's mind) is modeled as a latent variable and denoted as the unshaded node in Figure \ref{co-inmailed}. Given the information need, the searcher generates a keyword query in the Query-By-Keyword system and gets a list of results. The query and results are observed variables, denoted as shaded nodes. If the searcher decides to reach out to some of the results, these results are likely to be relevant to the hidden information need. It is worth mentioning that on our premium talent search product, searchers are charged some amount when they send messages, typically called \textit{Inmails}, to results. So, sending Inmail is strong evidence of result relevance. Since each of the inmailed results is a good fit for the hiring position, if (hypothetically) the searcher used \emph{Search By Ideal Candidates} instead of the Query-By-Keyword system, one or a few of these results could be used as ideal candidate(s) representing the information need. The rest of inmailed results (i.e., \textit{co-inmailed} results) could be considered as relevant ones and the results without inmail are non-relevant (Figure \ref{co-inmailed}).

\begin{figure}
\centering
\includegraphics[width=0.25\textwidth]{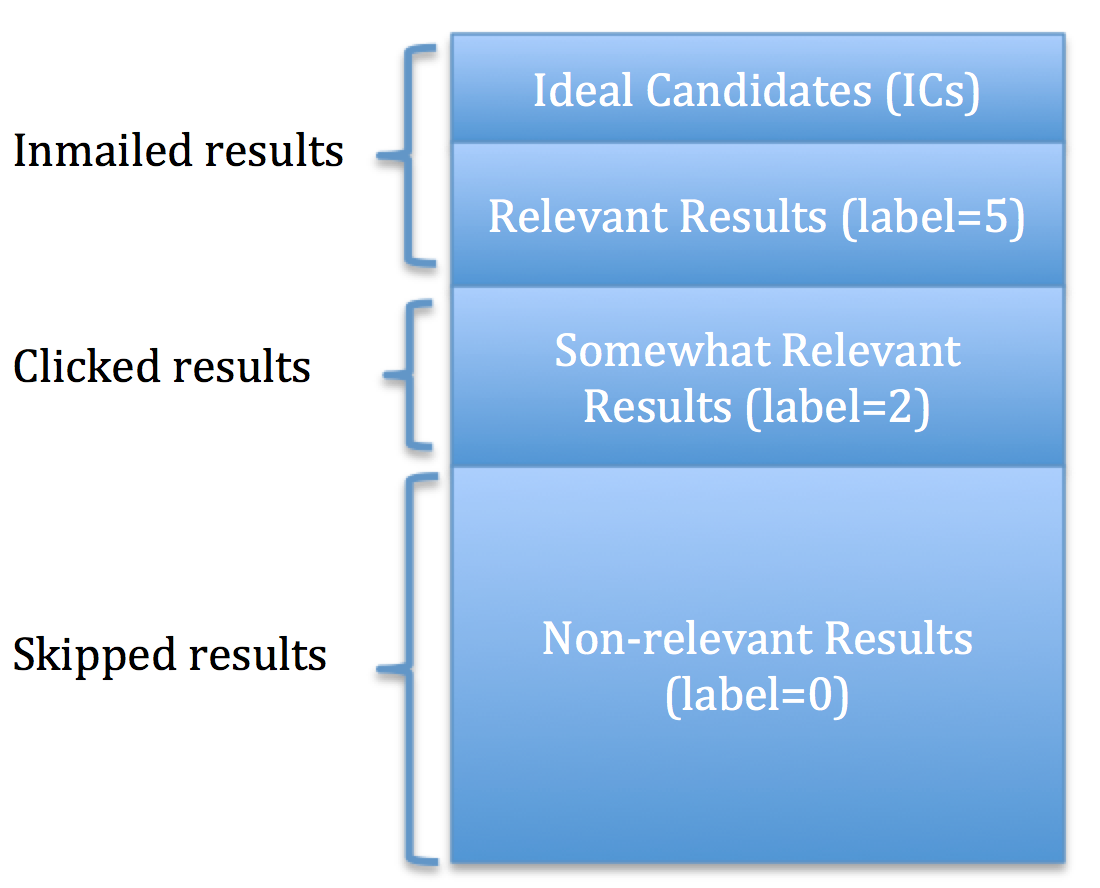}
\caption{Graded relevance labels derived from searcher's actions on results.}
\label{co-inmailed_2}
\end{figure}

Given the generative model, we define a labeling strategy as the following: within each search, we randomly pick one, two or three of the inmailed results as ideal candidates (Figure \ref{co-inmailed_2}). The rest of the results with inmail are considered to be relevant to the ideal candidates (thus assigned to the highest relevance label, e.g., 5). To make relevance labels more granular, amongst the results without inmails, the clicked results are considered somewhat relevant  (assigned to a mid-level label, e.g., 2). Finally, the skipped results are considered non-relevant (assigned to the lowest label, e.g., 0).

To avoid position bias in searchers' behaviors, we use a top-K randomized bucket on the Query-By-Keyword system, in which the top-K results are randomly shuffled (K=100), collect searchers' actions on the results and infer training data as described above.  When collecting the data, we count every query issued as a unique search. For instance, if the query ``software engineer'' is issued five times, they are considered as five distinct searches. The reason is that even though the query contents being searched for are identical across the five searches, they may have different search contexts when considering other dimensions such as searcher, location or time, etc. Moreover, this keeps the query distribution of the labeled data the same as of the live search traffics. We run the randomization bucket as mentioned above for two full weeks. The reason we keep the period in the order of weeks is to iron out strong weekly patterns. For example, if the labeled data is collected only during weekdays, it will be biased towards weekday search patterns but not to represent weekend search traffics. After two weeks, the randomization bucket ends up with about one hundred thousand searches with at least two Inmails from tens of thousands of unique searchers. The labeled data is then divided into training, validation and test sets.

\subsection{Proposed Features}
\label{proposed_features}
Typically, a search ranking function orders results by semantic similarity between each result and information need. In \emph{Search By Ideal Candidates}, the information need is represented by a query (generated by the query builder), searcher information as well as ideal candidates explicitly entered by the searcher. Given the characteristic, we propose a new set of features for \emph{Search By Ideal Candidates} that directly capture the resemblance between the ideal candidates and results on various aspects including expertise, title and career-path similarities.

\subsubsection{Expertise Similarity}
Skill section in a member profile showcases his or her professional expertise and is one of the most important parts in the profile. To capture the expertise similarity between ideal candidates and each result, we measure the similarity between their skill sets, and construct two features for this. The first one is \textit{Jaccard} similarity between the skill sets of each ideal candidate $c$ and each result $r$. If there are multiple ideal candidates, we take the average across them as in the equations below. 

\begin{gather*}
	JaccardSim(c, r) = \frac{|skills(c) \cap skills(r)|}{|skills(c) \cup skills(r)|} \\
	SkillJaccardSim(IC, r) = \frac{\sum_{c \in IC} {JaccardSim(c, r)}}{|IC|}
\label{skill_Jaccard}
\end{gather*}

As aforementioned, not every skill in a profile is equally important to the corresponding member. Indeed, some skills are more relevant to the core expertise than the others. To reflect this, in the second feature, we use \textit{Cosine} similarity between an ideal candidate and a result in which the skills are weighted by the expertise scores (denoted as $exSc$ in Equation \ref{skill_cosine}) described earlier in Section \ref{query_building_sec}.

\begin{equation}
	CosineSim(c, r) = \frac{\sum_{sk}{exSc(sk, c)exSc(sk, r)}}{\sqrt{\sum_{sk}{exSc(sk, c)^2}}\sqrt{\sum_{sk}{exSc(sk, r)^2}}}
\label{skill_cosine}
\end{equation}

\subsubsection{Title Similarity}
Besides skills, current job title is another crucial part in a profile that concisely describes job function of a member. To capture the job similarity, we compute Jaccard and Cosine similarity between the titles of each candidate and each result. For these features, we remove stop-words and treat a title as a bag of words. Similar to expertise similarity, in the case that a searcher enters multiple ideal candidates, we take the arithmetic average of all similarities as the final output.      

\subsubsection{Career-Path Similarity}
The expertise and title similarities described above are based on the \textit{accumulative} and \textit{current snapshots}, respectively, of ideal candidates and results. These features do not reflect \textit{temporal} career trajectories of them. Thus, we propose career-path similarity feature ascertaining similarity between two members by considering their temporal sequences of job positions. Specifically, each member is modeled by a sequence of nodes, each of which records all information within a particular position the member holds. At the node level, we use a supervised model combining various signals capturing the similarity between two  positions. Bellows are some important features.

\begin{itemize}
  \item \textit{Company match:} whether or not the two positions are at the same company.
  \item \textit{Company similarity:} if the two positions are at different companies, we consider similarity between the companies. For instance, LinkedIn is more similar to Facebook than Shell. This is estimated by co-viewing relationship, e.g., people viewing LinkedIn company page also view Facebook.
  \item \textit{Industry match:} whether or not the two positions belong to the same industry. Some examples of industries include Internet and Human Resource.
  \item \textit{Industry similarity:} since each company usually belongs to an industry, industry similarity can also be derived from the company browse map described above.  
  \item \textit{Title match:} whether or not the two positions have the same title.
  \item \textit{Title similarity:} for instance, ``Software Engineer'' and ``Software Developer'' are similar titles.
  \item \textit{Position description similarity:} this feature extracts keywords from the two position descriptions then computes cosine similarity between the two. 
  \item \textit{Seniority similarity:} each position is mapped to a seniority level, e.g., ``Senior Director of Engineering'' is more senior than ``Director of Engineering''. Then, we compute the similarity based on the levels.
\end{itemize}

Given two career paths $P=[p_1, p_2 ... p_n]$ and $Q=[q_1, q_2 ... q_m]$, the distance between them $D^{seq}$ is the sum of node pair distances in the optimal alignment between the two sequences (here we use distance and similarity interchangeably since one is just the reserve of the other). The optimal alignment could be determined by a dynamic programming algorithm based on Equation \ref{sequence_alignment}. In this equation, parameter $\lambda$ is a \textit{gap penalty} that allows dissimilar nodes to be skipped with some penalty for the non-contiguity.

\begin{equation}
	D^{seq}(P^i, Q^j) = \\
	                   max \begin{cases}
	                           D^{seq}(P^{(i-1)}, Q^{(j-1)})+D^{node}(p_i, q_j) \\
	                           D^{seq}(P^{(i-1)}, Q^j)+\lambda \\
	                           D^{seq}(P^i, Q^{(j-1)})+\lambda
                            \end{cases}
\label{sequence_alignment}
\end{equation}

\subsection{Existing Features and Learning Algorithm}
\label{existing_features}
Besides the new features designed specifically for Query-By-Example paradigm, we also use standard features used in Query-By-Keyword talent search systems. These existing features are generally divided into the following categories.

\textbf{Skill expertise feature:}
one of the most important signals on talent search is to match each result's expertise with the required one \cite{hathuc2015expertisesearch}. To capture the match, we use a skill expertise feature, which is a sum of each result's expertise scores on the skills in the query. 

\textbf{Text matching features:} 
the most traditional type of features in information retrieval is textual features. These features match the keywords in queries with different sections of each result's profile, such as, title, company, etc. 

\textbf{Entity-aware matching features:}
for this kind of features, we first semantically tag each query with entity types, such as title, company and skill. Then,  match the keywords with the corresponding field in results (member profiles).

\textbf{Result prior scores:}
these features are query-independent. They are designed to capture result quality, such as spam score, profile quality, historical CTR (popularity), etc.  

\textbf{Geographic features (personalized features):} talent search on LinkedIn is highly personalized. For instance, a query like ``software developer" will produce very different results if the searcher is in New York City, USA as opposed to (say) Perth, Australia. Location plays an important role in personalizing the results.

\textbf{Social features (personalized features):} another important aspect of personalization is to capture how the results socially relate to the searcher. We leverage a variety of the signals on LinkedIn, such as, how the searcher socially connects with the company posting the job, e.g., if he or she follows the company or the searcher has friends working at the company, etc. to generate the features in this category. We refer readers to \cite{hathuc2015expertisesearch, Li16} for a more detailed description of the existing features.

Given the training data and the features, we apply Coordinate Ascent \cite{metzler2007linear}, a listwise learning-to-rank algorithm, to search for an optimal model. For efficiency purpose, we use linear models in our work. Coordinate Ascent algorithm has also been shown to be effective for learning linear ranking functions in some other search domains \cite{metzler2007linear}. One key benefit of listwise learning-to-rank approach over pointwise and pairwise ones is that the listwise approach can optimize ranking-based metrics directly \cite{Li2011, Liu2010}. The objective function we optimize in the learning process is normalized discounted cumulative gain (NDCG@K) defined on the graded relevance labels as described above. Parameter $K$ is set to \textit{15}, which is the average rank of the last seen/interacted results in the search result pages.

\section{Experiments}
\label{experiments}
\subsection{Skill Selection Evaluation}
Evaluating query building strategies is challenging because ground truth data is difficult to collect. A typical way to get ground truth data is human judgment. Given a set of ideal candidate profiles and alternative generated queries, annotators are asked to judge which queries can lead to good search results with respect to the ideal candidates. However, this is a complicated task even for professional recruiters. Moreover, given the highly diverse nature of LinkedIn members that are from many companies, countries and industries with various skill sets, getting human annotation for a collection of ideal candidate queries with decent coverage is expensive.

In this work, we leverage the Co-Inmail data (presented in Section \ref{training_data_generation}) to evaluate the skill selection strategy described in Section \ref{query_building_sec}. Given a set of ideal candidates $IC$ (randomly sampled from the results with Inmails), relevant results $R^+$ (the Co-Inmailed results) and non-relevant results $R^-$ (the ones without Inmails), we evaluate the skills selected by a query builder $S =  \{sk_1, sk_2 ... sk_K\}$ from the candidates. These skills are good ones to represent the expertise of the ideal candidates if they put the relevant results higher than non-relevant ones. In other words, the set $S$ is considered to be good if relevant results have higher expertise scores on these skills than non-relevant results. This can formally be described in the inequality below. 

\begin{gather*}
	AvgExpertise(R^+, S) > AvgExpertise(R^-, S) \\
	AvgExpertise(R^*, S) = \frac{\sum_{sk_i \in S, r \in R^*}{exSc(sk_i, r)}}{|R^*|}
\label{Skill_selection_eval}
\end{gather*}

Accuracy of a skill selection strategy is then the percentage of the searches in Co-Inmail data that satisfies the inequality above, i.e., correctly differentiates relevant results vs. non-relevant ones. We compare the skill selection approach in Section \ref{query_building_sec} that picks Top-K skills with a \textit{baseline} of randomly picking $K=10$ explicit skills from the ICs' profiles (denoted as Rand-10).

\begin{table}
\centering
\begin{tabular}{|l|c|c|c|c|} \hline
 & Rand-10 & Top-5 & Top-10 & Top-15 \\ \hline
Accuracy & 0.591  & 0.643 & 0.645 & 0.638\\ \hline
Improvement & - & +8.8\% & +9.1\% & +8.0\% \\ \hline
\end{tabular}
\caption{Skill Selection Performance}
\label{skill_selection_performance}
\end{table}

Accuracy of different skill selection strategies are shown in Table \ref{skill_selection_performance}. First, it is worth noting that even the baseline achieves accuracy well above $50\%$. That means randomly selected skills from the profiles of ideal candidates can still separate relevant results from non-relevant ones. More specifically, these skills are closer to the expertise of relevant results than the expertise of non-relevant ones. When the skills are selected by expertise scores of the ideal candidates on them, the accuracies, with the values of $K$ equals to 5, 10 and 15, are $8.8\%$, $9.1\%$ and $8.0\%$ higher than the baseline, respectively. That demonstrates the effectiveness on the expertise scores for skill selection task.

An alternative way to evaluate query building is to conduct A/B tests on various strategies on live traffic. Then, evaluate each of them based on searchers' interactions with generated queries and overall search performance. We leave this direction for future work.

\subsection{Search Ranking Evaluation}
\subsubsection{Feature Analysis}
To understand feature effectiveness, we compute Pearson correlation coefficient between feature values and labels of the corresponding results within each search (list), then take the average across the searches in Co-Inmail training set. The average correlation is used to evaluate features. Table \ref{top_ranking_features} shows five most important features. Interestingly, the first three are the features directly capturing the similarity between ideal candidates and results. That confirms the importance of the proposed features in \emph{Search By Ideal Candidates}. Between the two skill similarity features, Cosine similarity is ranked higher. That shows the benefit of weighting skills by expertise scores. Finally, the next two features are existing features including Entity-Aware Title Match (matching query terms tagged as a title with title field in each result) and result historical CTRs (a proxy of result popularity). This is inline with our observation from past Query-By-Keyword experiments. 

\begin{table}
\centering
\begin{tabular}{|l|l|} \hline
Rank & Feature \\ \hline
1 & Career Path Similarity \\ \hline
2 & Cosine Skill Similarity \\ \hline
3 & Jaccard Skill Similarity \\ \hline
4 & Entity-Aware Title Match  \\ \hline
5 & Result Historical CTR \\ \hline
\end{tabular}
\caption{Top Ranking Features}
\label{top_ranking_features}
\end{table}

\subsubsection{Ranking Performance on Co-Inmail Data}
After training a ranking model for \emph{Search By Ideal Candidates} using all new features as well as existing ones, we evaluate the model on a held-out test set, by comparing the model with two baselines. The \textit{first baseline} is the current model in production for the Query-By-Keyword system. This model uses all existing features described in Section \ref{existing_features}. The \textit{second baseline} also uses all existing features, but is trained on the Co-Inmail training data. This baseline does not use the new features in Section \ref{proposed_features}.

\begin{table}
\centering
\begin{tabular}{|l|c|c|c|} \hline
 & NDCG@5 & NDCG@15 & NDCG@25 \\ \hline
Over baseline 1 & +16.9\%  & +12\% & +9.9\% \\ \hline
Over baseline 2 & +6.7\% & +5.3\% & +4.4\% \\ \hline
\end{tabular}
\caption{Ranking performance on Co-Inmail data. All of the improvements are statistically significant with $p-value  < 0.01$ using the Student's paired t-test.}
\label{ranking_performance_coinmail}
\end{table}

Table \ref{ranking_performance_coinmail} presents relative improvements of the new ranking model compared to the two baselines, respectively. We use Normalized Discounted Cumulative Gain at K (NDCG@K) metrics, where $K$ equals to 5 (focus on the top results), 15 (the average rank of the last seen/interacted results) and 25 (the number of results on the first page). Between the two baselines, even though using the same features, baseline 2 which is trained for Query-By-Example paradigm yields better results across the metrics (i.e., the improvements of the new model over baseline 2 are smaller than the improvements over baseline 1). This demonstrates that it is important to train ranking models specifically for Query-By-Example paradigm. The reason is that the queries extracted from ideal candidates (which typically have 10 skills + 10 companies + several titles) are different from queries entered by searchers in the traditional Query-By-Keyword system (which usually contain from 5 to 10 keywords).

Focusing on the new ranking model, it is significantly better than both baselines across all metrics. For instance, on NDCG@5, the new model is $16.9\%$ and $6.7\%$ better than baselines 1 and 2, respectively. The difference between the new model and baseline 2 (both are trained on the same training data) confirms the effectiveness of the proposed features that are specifically designed for Query-By-Example paradigm.

\subsubsection{Ranking Performance on Randomized Data}
\label{ranking_performane_on_randomized_data}
Post launch of the new \emph{Search By Ideal Candidates} system, in retrospect, we re-evaluate these three models in a more direct way. Instead of generating the test set from Co-Inmail data, we launch a randomized bucket in a small random fraction of live traffics of the \emph{Search By Ideal Candidates} system for two weeks. In this bucket, top K ($K=100$) results are randomly shuffled before showing to searchers. Our previous experiments in the context of LinkedIn search suggest that with K sufficiently large (e.g., 100), offline experiment results on top-K randomized data are directionally inline with online A/B test results \cite{Li16,hathuc2015expertisesearch}.  We collect searchers' actions on the results and assign labels to them in the same way as before. Since this bucket is collected directly from the product usage, this dataset is superior in regard to model evaluation. The final test set includes more than two thousand and five hundred searches from hundreds of unique searchers.

Given the randomized data, we evaluate the models using the same metrics as shown in Table \ref{ranking_performance_randomized}. Similar results between Table \ref{ranking_performance_coinmail} and Table \ref{ranking_performance_randomized} (baseline 1 < baseline 2 < the new model) show an agreement between Co-Inmail data and the randomized data, which confirms the merit the Co-Inmail approach that infers training data for \emph{Search By Ideal Candidates} from usage logs of the previous Query-By-Keyword system. 

Compared with the baselines, the new model is also significantly better. For instance, the new model is $6.2\%$, $4.4\%$ and $3.4\%$ better than baseline \textit{2} in terms of NDCG@5, NDCG@15 and NDCG@25. These gains are attributed to the new features. It is worth re-emphasizing that in our design, the retrieval and ranking steps are dependent on the generated queries. While the queries are essential to the searchers in terms of understanding and controlling over the results, translating the ideal candidates to the queries is likely to result in some information loss since the queries cannot perfectly represent the candidates (this is a design trade-off). Thus, another way to view the proposed features is that they present a way to recover the information loss since they capture the similarities between results and the input candidates directly. The difference between our model and baseline \textit{2} again re-confirms the importance of them.

\begin{table}
\centering
\begin{tabular}{|l|c|c|c|} \hline
 & NDCG@5 & NDCG@15 & NDCG@25 \\ \hline
Over baseline 1 & +13.1\%  & +8.4\% &  +6.5\%\\ \hline
Over baseline 2 & +6.2\% & +4.4\% & +3.4\% \\ \hline
\end{tabular}
\caption{Ranking performance on Randomized data. All of the improvements are statistically significant with $p-value  < 0.01$ using the Student's paired t-test.}
\label{ranking_performance_randomized}
\end{table}

\subsection{Error Analysis}
To gain further insight on the system performance, we conduct an error analysis on the cases where the end-to-end system does not work well. Specifically, we randomly sample 100 searches from 100 unique searchers where they abandon the searches, i.e., have no actions on the results. Then, we manually identify the failure reasons.

In about two thirds of the cases, the searches are abandoned because of the generated queries. A deep dive into these queries and their corresponding ideal candidates suggests that the most frequent root cause of this is because the ideal candidates use non-standardized job titles, such as, ``ethical iOS hacker'', ``senior rocket scientist'' or ``lead data werewolf''. These titles are infrequent, thus they match few results. Moreover, since they cannot be mapped to standardized titles in the database, the system fails to expand them to similar titles. As a result, there are few or even none results returned for these queries (recall that we take a conjunction across attribute types in each query, as described in Section \ref{query_building_sec}). Moving forward, a solution for this kind of errors is to improve the coverage of the titles that can be standardized. Moreover, in the case that all of the titles extracted from the ideal candidates are non-standardized, we can experiment with the idea of totally ignoring this attribute type in queries in the retrieval phase.

In the rest of the cases, the problem is on result rankings. A common pattern in this category is that the input ideal candidate is an engineer from a \textit{big tech company} and also happens to be the founder (or CEO/CTO, etc.) of a \textit{startup}. Thus, the generated query would be: \textit{Q = title:(``engineer'' OR ``founder'') AND company:(big company OR startup) ...} In this pattern, occasionally the ranker mistakenly ranks the founder (or CEO/CTO, etc.) of the \textit{big company} on top because the result well matches the query. The root cause of this error is that after the queries are generated, the associations between entities across attribute types (e.g., title and company of the same position of an ideal candidate) are lost in the queries. The career-path similarity feature directly capturing similarities between ideal candidates and results is designed exactly to recover for the loss. However, in some cases, it is out-weighted by many query-document matching features. This is left as a future direction.

\section{Conclusions}
\label{conclusions}
This paper introduces the next generation of talent search at LinkedIn, which drives about {\it 64\%} of the company revenue. Throughout the paper, we present a comprehensive solution for the problem of how to transition from Query-By-Keyword to Query-By-Example for talent search. In particular, we detail how to resolve practical challenges when building such system in an industrial setting, including: imperfection of profile data (e.g., outlier and missing skills), the lack of personalized training data when transitioning to a new search paradigm and how to customize ranking features for Query-By-Example. 

We summarize the major design choices, tradeoffs and the lessons learned throughout the course of this work:
\begin{itemize}
\item Given that the biggest challenge for our users is how to translate the criteria of a hiring position into a search query, we design a new system allowing searchers to just input one or a few ideal candidates instead. In most of the cases, they can simply pick existing team members, thus this design significantly reduces their efforts.
\item In recruiting process, system transparency and the ability to control on results are essential to users. Thus, we decide not to take a ``black box'' item-to-item recommendation approach. We instead approach this as a search problem by explicitly generating queries and allowing searchers to modify them. Even though this approach would limit some relevance modeling power (as analyzed in Section \ref{ranking_performane_on_randomized_data}), it achieves a better balance between system transparency and user control vs. relevance.
\item To alleviate the limit, we propose an additional set of features capturing the similarities between results and the input candidates directly, thus recovering the information loss in the process of translating the candidates into queries.
\item Another lesson is for highly complicated signals like skill expertise scores and career path similarity, a two-phase architecture including an offline and an online phase achieves a balanced tradeoff between effectiveness and efficiency. The offline phase runs on distributed computing platforms like Hadoop, thus allows processing very large datasets with complex methods. It periodically generates new versions of the signals offline. The online phase then simply consumes the latest version of the signals in real time.
\end{itemize}

Our experiments confirm the effectiveness of the proposed approach. Specifically, on skill selection, our approach is $9.1\%$ more accurate than the baseline. On search ranking, the new ranking model is significantly better than the other two baselines across all metrics on both Co-Inmail and randomized test sets. That demonstrates the value of the training data generation approach as well as the new features. Currently, the query building approach and the search ranking model serve all live traffics of \emph{Search By Ideal Candidates} at LinkedIn. Although we focus on talent search in this paper, we believe that these lessons are also useful when moving from Query-By-Keyword to Query-By-Example in other domains.

\textbf{ACKNOWLEDGMENT:} We would like to thank Ye Xu, Satya Pradeep Kanduri, Shan Zhou and Sunil Nagaraj for their contributions during the course of this work.

\bibliographystyle{abbrv}
\bibliography{sigproc} 

\end{document}